\documentclass[aps,pre,groupedaddress,twocolumn]{revtex4}

\usepackage[pdftex]{graphicx}
\usepackage{epstopdf}
\usepackage{amsmath}
\usepackage{natbib}

\begin{document}

\author{\'E. Rold\'an$^{1\dagger}$, I. A. Mart\'inez$^{2\dagger}$,  J. M. R. Parrondo$^{1}$\footnote{Corresponding author: \url{parrondo@fis.ucm.es}}, D. Petrov$^{2,3}$\medskip}

\address{$^1$Departamento de F\'isica At\'omica, Molecular y Nuclear and GISC, Universidad Complutense de Madrid, 28040,  Madrid, Spain\\
$^2$ICFO $-$ Institut de Ci\`encies Fot\`oniques, Mediterranean Technology Park,  Av. Carl Friedrich Gauss, 3, 08860, Castelldefels (Barcelona), Spain \\
$^3$ICREA $-$ Instituci\'o Catalana de Recerca i Estudis Avan\c cats,\\Passeig Llu\'is Companys, 23, 08010, Barcelona, Spain\\
\medskip $^{\dagger}$ These authors contributed equally to this work}

\title{Universal features in the energetics of symmetry breaking}

\maketitle

{\bf  A symmetry breaking (SB) involves an abrupt change in the set of microstates that a system can explore. 
This change has unavoidable thermodynamic implications. 
According to Boltzmann's microscopic interpretation of entropy, a shrinkage of the 
set of compatible states implies a decrease of entropy, which eventually needs to be compensated by dissipation of heat and consequently requires work.  Examples are the compression of a gas and the erasure of information. On the other hand, in a spontaneous SB, the available phase space volume changes without the need for work, yielding an apparent decrease of entropy. Here we show that this decrease of entropy is a key ingredient in the Szilard engine and Landauer's principle and report on a direct measurement of the entropy change along SB transitions in a Brownian particle. The SB is induced by a bistable potential created with two optical traps. The experiment confirms  theoretical results based on fluctuation theorems, allows us to reproduce the Szilard engine extracting energy from a single thermal bath, and shows that the signature of a SB in the energetics is measurable, providing new methods to detect, for example, the coexistence of metastable states in macromolecules.}

\vspace{0.1cm}

When a symmetry is broken, a system ``makes a choice'' from among a set of instances $i=1,\dots,m$. For a classical infinite system, symmetry breaking (SB) consists of  a sudden change in the set of available states:  the whole phase space $\Gamma$ is partitioned into non-overlapping regions $\Gamma_i$, corresponding to the different  instances $i=1,\dots,m$. 
The partition occurs when a certain control parameter $\lambda$ crosses a critical value $\lambda_c$ above which the system can no longer move spontaneously from one region to another and gets confined within $\Gamma_i$ with  probability $p_i$, $\sum_ip_i=1$. The notion of SB can be extended to finite systems with metastable states. The confinement is not strict in this case: the system can jump form a region $\Gamma_i$ to another $\Gamma_j$. However,  if the average residence time in each region is much larger that the time scale of the process under consideration, one can talk about an effective SB. In this case, the SB transition is not localized at a single value of the control parameter $\lambda$,  but is  rather a continuous transition where  metastable states develop. 

The energetics associated to SB transitions and, in general, to the manipulation of metastable states has special relevance to a number of interesting physical situations, some of them  realized experimentally in the last years. The original Szilard engine, a refined version of the original Maxwell demon, can extract work from a single thermal bath  using the information created in a SB \cite{leff1990maxwell,parrondo2001szilard,marathe2010cooling,toyabe2010experimental}. Landauer's principle accounts for the minimum dissipation associated to the erasure of information, which  is a manipulation of the two metastable states making up  a single bit memory \cite{leff1990maxwell}. The erasure can be interpreted as the restoration of a broken symmetry (see below) and has been reproduced with a Brownian particle in a double well potential created by optical tweezers \cite{berut2012experimental}. In molecules, metastable states correspond to different molecular conformations as well as to  kinetic states of special relevance to biophysics. The energetics of processes involving metastable states has become a tool to measure conformational free energies in those contexts \cite{maragakis2008differential,junier2009recovery,alemany2012experimental}. An extended version of the non-equilibrium work theorem relates the probability distribution of the work in a process connecting two metastable states with their conformational free energies. Maragakis et al. \cite{maragakis2008differential} applied this result to numerical simulations of switches between two different conformations of alanine dipeptide. In \cite{junier2009recovery,alemany2012experimental}, this generalized work theorem is used to estimate, from stretching experiments, the conformational free energy of DNA hairpins that possess intermediate or misfolded kinetic states.

In this paper, we report on an experimental realization of a SB consisting of a continuous transition from a single well to a double well potential affecting a Brownian particle. We reproduce the transition by moving apart two optical traps and then measure the heat dissipated by the particle to the surrounding water that acts as a thermal reservoir. An electrostatic field acting on electrical charges on the particle's surface allows us to tune the bias towards one or the other trap and explore the relation between the energetics of the SB and the probability of adopting one of the instances. We finally build a Szilard engine as a SB followed by the restoration of that symmetry under different conditions. This process completes a cycle that extracts energy from the thermal bath if the electrostatic field along the process is properly chosen.

\section{Symmetry breaking and symmetry restoration}

Consider a system with Hamiltonian   $\mathcal{H}(x;\lambda)$ ($x\in\Gamma$), depending on a control parameter $\lambda$, and an isothermal process at temperature $T$ involving a SB, where the parameter changes in time as $\lambda(t)$ with $t\in [t_{\rm ini},t_{\rm fin}]$. The average work required to complete the process, when the system adopts instance $i$, is bound by
\begin{equation}\label{dissipativework}
\langle W\rangle^{\rm (SB)}_i-\Delta F_i \geq kT\ln p_i,
\end{equation}
where $k$ is the Boltzmann constant and $\Delta F_i=F_{{\rm fin},i}-F_{\rm ini}$ is the change in free energy. The initial free energy is defined as usual, $F_{\rm ini}=-kT\ln Z(T,\lambda(t_{\rm ini}))$ where  $Z(T,\lambda)=\int_\Gamma dx\,e^{-\beta \mathcal{H}(x;\lambda)}$ is the partition function of the system. On the other hand, the final free energy $F_{{\rm fin},i}=-kT\ln Z_i(T,\lambda(t_{\rm fin}))$ is a conformational free energy defined in terms of the partition function restricted to the region $\Gamma_i$, i.e., $Z_i(T,\lambda)=\int_{\Gamma_i} dx\,e^{-\beta \mathcal{H}(x;\lambda)}$. The bound in Equation \eqref{dissipativework} is met with equality if the process is quasistatic. Recalling the relation between the free energy, $F$, the internal energy $E$, and the entropy $S$ of a system, $F=E-TS$, and the First Law of Thermodynamics $\Delta E=W+Q$, where $Q$ is the heat or energy transfer from the thermal reservoir to the system, we easily derive a bound for the conformational entropy production:
\begin{equation}\label{mainS}
\langle S_{\rm prod}\rangle_i^{\rm (SB)}\equiv \Delta S_i -\frac{\langle Q\rangle^{\rm (SB)}_i}{T}\geq k\ln p_i .
\end{equation}

A rigorous proof of these bounds follows from fluctuation theorems (see Supplementary Information). However, the origin of the term $k\ln p_i$ in Eqs.~(\ref{dissipativework}) and (\ref{mainS}) can be easily understood. A SB comprises a contraction of the set of available states from $\Gamma$ to $\Gamma_i$ without the need for any extra work \cite{parrondo2001szilard,marathe2010cooling}. This amounts to an increase in free energy  $-kT\ln (Z_i/Z)$ which is not compensated by work and heat dissipation. Assuming an instantaneous SB, $p_i=Z_i/Z$, yielding  the extra term $kT\ln p_i$ in Eqs.~(\ref{dissipativework}) and (\ref{mainS}). 

This work-free shrinkage of the available phase space is entirely due to the SB transition and is not in contradiction with the Second Law of Thermodynamics, because the final state $\rho_{i}(x)$ is not in complete equilibrium and  the final entropy cannot be considered as a true thermodynamic entropy.  In some contexts, $S_i$ and  $F_i$  are called, respectively, the conformational entropy and the conformational free energy, but they are not true thermodynamic potentials (they are not state functions, for example \cite{horowitz2013optimizing}). However, both are useful tools for analyzing the energetics of processes involving SB transitions \cite{sagawa2009minimal,horowitz2012imitating,horowitz2013optimizing}. An alternative interpretation of the compatibility between the Second Law and the decrease of entropy in Eq.~(\ref{mainS}) is that the latter is compensated by an increase of the meso- or macro-scopic uncertainty, quantified by the Shannon entropy of the SB outcome, $H(p_i)=-\sum_i p_i\ln p_i$. Notice that the average of Eq.~(\ref{dissipativework}) over $p_i$  yields precisely $kT\,H(p_i)$.

Similar inequalities hold for  a process where a symmetry is restored. To assess the energetics of a symmetry restoration (SR) we have to consider the time reversal of the restoration, which is a SB. Let us call $\tilde p_i$ the probability that the system adopts instance $i$ in this SB resulting from the time reversal of the original process. Under  time reversal, the reversible work and the increase in the free energy change sign. Therefore (see Supplementary Information for a detailed proof):
\begin{equation}\label{main2}
\langle W\rangle^{\rm (SR)}_i -\Delta F_i   \ge -kT\ln \tilde p_i,
\end{equation}
where now $\Delta F_{i}=F_{\rm fin}-F_{{\rm ini},i}$ is the free energy change of the SR. Notice that now it is the initial free energy that depends on the instance $i$. For the entropy:
\begin{equation}\label{mainS2}
\langle  S_{\rm prod} \rangle^{\rm (SR)}_i   \ge -k\ln \tilde p_i.
\end{equation}
The aim of this paper is to check experimentally Eqs.~\eqref{mainS} and \eqref{mainS2}, which have important implications in the thermodynamics of information processing and the foundations of statistical mechanics. For instance, Landauer's principle follows immediately from Eq.~\eqref{main2} applied to a one-bit memory consisting of a physical system with two stable states, 0 and 1, each one with the same free energy $F_0=F_1$. The minimal cost of erasing a  bit or, more precisely, to drive bit $i=0$ or $1$ to the state 0 (restore-to-zero operation) is $\langle W\rangle^{\rm erasure}_i\ge -kT\ln \tilde p_i+\Delta F_i=-kT\ln \tilde p_i$
for $i=0$ or 1, since in both cases $\Delta F_i=F_0-F_i=0$. If the initial bit is unknown, the best we can do is $\tilde p_i=1/2$ and $\langle W\rangle^{\rm erasure}_i\ge kT\ln 2$ \cite{leff1990maxwell,berut2012experimental}.

The energetics of the Szilard engine \cite{sagawa2009minimal,horowitz2010nonequilibrium,horowitz2011thermodynamic,horowitz2011iopscience} can be as well  easily reproduced from Eqs.~\eqref{dissipativework} and \eqref{main2}. In the Szilard setup, a system  undergoes a SB and chooses between two instances 0 or 1 with probability $p_0$ and $p_1$, respectively. Then we measure the instance that has been chosen and restore the broken symmetry  driving the system back to the original state through some protocol $\lambda_i(t)$. The time reversal of this protocol is a SB transition with possibly different probabilities $\tilde p^i_j$. Notice that the superscript $i$ refers to the protocol $\lambda_i(t)$ implemented when $i$ is measured, whereas the subscript $j$ refers to the probability of obtaining outcome $j$ if the protocol $\lambda_i(t)$ is reversed \cite{horowitz2013optimizing,horowitz2011iopscience}. The work necessary to implement the SB is bound by Eq.~\eqref{dissipativework} and the work necessary to restore the symmetry is  bound by  Eq.~\eqref{main2}. Therefore, the  total average work that we have to perform to run the whole cycle obeys
\begin{eqnarray}
\langle W\rangle &=& \sum_i p_i\left[\langle W\rangle^{\rm (SB)}_i+\langle W\rangle^{\rm (SR)}_i\right] \nonumber\\
 &\ge & kT\sum_i p_i\ln\frac{p_i}{\tilde p^i_i},
 \label{work}
\end{eqnarray}
and $\langle S_{\rm prod}\rangle=\langle W\rangle/T$.
In the case of the original Szilard engine, $p_i=1/2$ and $\tilde p_i^i=1$, yielding $\langle W\rangle \ge - kT\ln 2$, i.e., the extraction of an energy $kT\ln 2$ in a cycle. If the demon does not use information from the measurement performing always the same protocol, i.e., $\lambda_i(t)=\lambda(t)$, then $\tilde p^i_j=\tilde p_j$ normalized to unity $\sum_i \tilde p_i=1$ yielding 
$\langle W\rangle =  kTD( p_i||\tilde p_i)\ge 0$, where $D(p||q)$ is the relative entropy between the two probability distributions $p$ and $q$ \cite{horowitz2013optimizing,kawai,parrnjp}. To build a Szilard engine, it is enough to find $p_i$ and $\tilde p_i^i$ such that the average work $\langle W\rangle$ in Eq.~\eqref{work} is negative; for instance by choosing protocols where $\tilde p_i^i>p_i$ (see below for an explicit construction of the  engine and the Supplementary Information for an illustration of Eq.~(\ref{work})).

\section{Experimental test}

Inequalities~\eqref{mainS} and~\eqref{mainS2} are universal, i.e., they do not depend on the details of the SB or even on the physical nature of the system under consideration.  We have tested both inequalities experimentally using a Brownian particle in an optical trapping potential with a time-dependent profile.

\begin{figure}
\includegraphics[width=6cm]{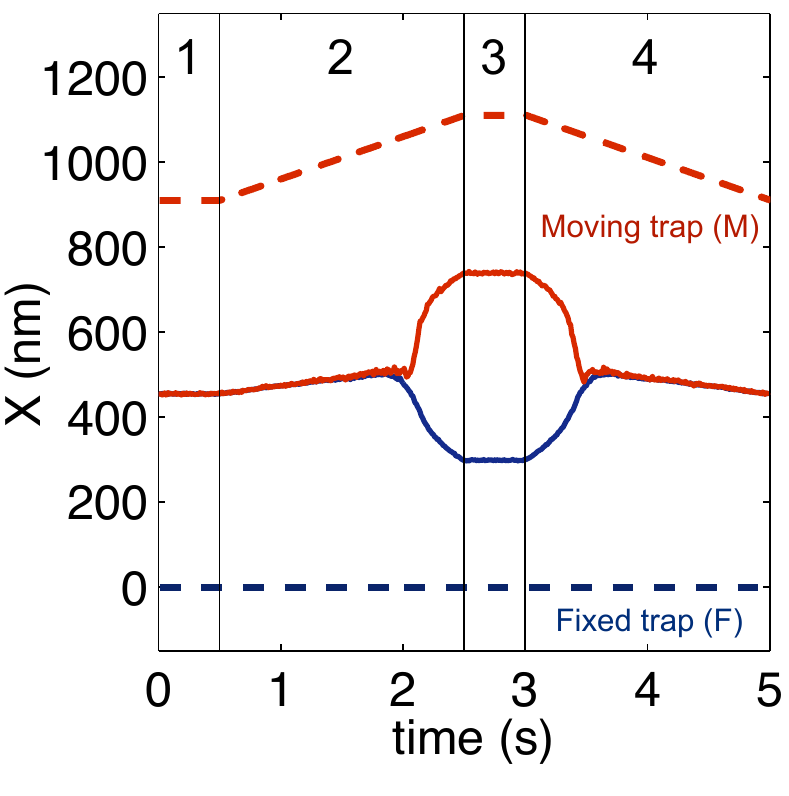}
\includegraphics[width=8.5cm]{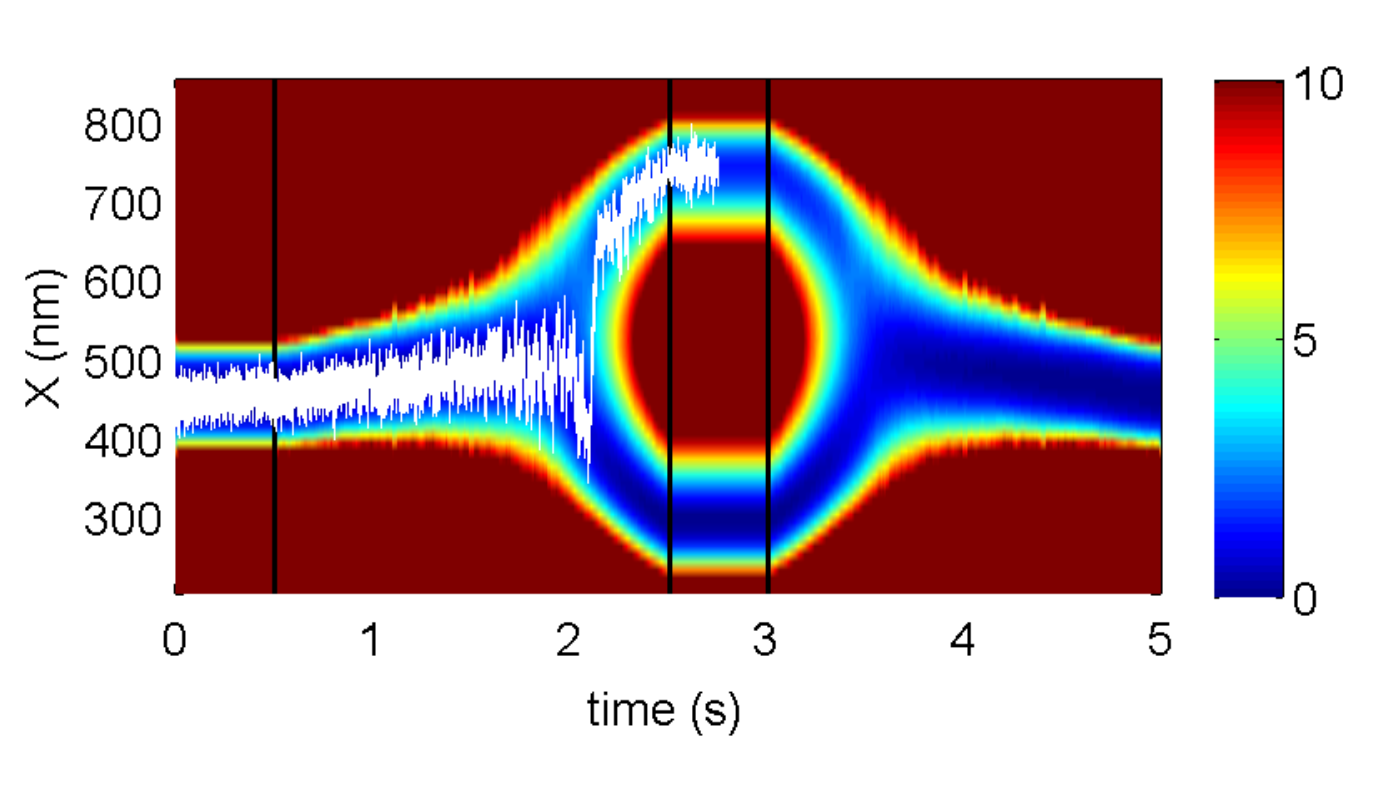}
\caption{Experimental protocol of symmetry breaking and symmetry restoration. Top. Positions of the $F$ trap (blue dashed line) and $M$ trap (red dashed line) as functions of time during the  protocol. 
 Ensemble average position of the trapped bead after 
implementing the protocol cyclically for $t = 2400 \rm \,s$ over $F$ trajectories (blue solid line) and $M$ trajectories (red solid line). 
Bottom. Spatial--temporal mapping of the  potential $U(x,t)$  obtained from the statistics of trajectories of the 
bead for $t= 2400 \rm \,s$ in the presence of an external force such that $p_{F}=0.8$. Color bar on the right indicates the depth of the potential energy (in units of $kT$). A single trajectory of the bead when it chooses the $M$ trap is also plotted (white line). }
\label{figcycle}
\end{figure}
We study the motion of a polystyrene spherical bead ($1\,\mu \rm m$ diameter) suspended in water in the presence of two optical traps (see Methods). One of the traps, labelled $F$, is held fixed at $x=0$ (Fig.~\ref{figcycle}, top panel). The other trap, labelled $M$, is moved along the $x-$axis following the four step protocol depicted in the top panel in Fig.~\ref{figcycle}. 
Initially the two traps with their centers separated by a distance 
$L_{\rm ini}=910\, \rm nm$ are at rest for a period of time $\tau_1=0.5\,\rm s$; (step 1).  Then the trap $M$ is moved along the $x-$axis at constant velocity $v_{\rm trap}$  for a time $\tau_2$ (step 2). During  step 3, the two traps with their centers separated by $L_{\rm fin}=1110\, \rm nm$
are again kept fixed for $\tau_3=0.5\, \rm s$. Finally, the trap $M$ is moved back from 
$L_{\rm fin}$ to its initial position  $L_{\rm ini}$ with velocity  $-v_{\rm trap}$ for a time $\tau_4=\tau_2$ (step 4). The total duration of the cycle is $\tau=\sum_{i=1}^4 \tau_i= 2\tau_2+1\,{\rm s}$.
By cyclically repeating this protocol, we 
can study both the SB (steps $1-2-3$) and the SR (steps $3-4-1$).

Due to the presence of inherited electrical charges at the surface of the bead, we can bias the motion of the bead towards the $M$ or $F$ trap by applying a voltage to electrodes inserted in the fluid chamber  \cite{Martinez2013} (see Supplementary Information). 

The protocol can be considered quasistatic for velocities around 100 nm/s or lower, for which the heat dissipation due to friction force is  on the order of $\gamma v_{\rm trap}^2\approx 10^{-22}\,{\rm J/s}\approx 0.02\,kT/{\rm s}$, where $\gamma= 6\pi R\eta$ is the friction coefficient, $R=0.5\,\mu{\rm m}$ is the radius of the bead, and $
\eta=8.9\times 10^{-4} \, \rm Pa\cdot s$ the dynamic viscosity of water at 25$^{\rm o}$C. We have implemented two quasistatic protocols with $v_{\rm trap}=100\,\rm nm/s$, $\tau_2=2\,\rm s$, and $v_{\rm trap}=36.36\,\rm nm/s$, $\tau_2=5.5\,\rm s$. 

During step 2,  Kramers transitions trigger the SB. This can be seen clearly in the trajectory of the bead  presented in the bottom panel of Fig. \ref{figcycle}. At the end of the SB protocol (steps $1-2-3$), Kramers transitions are not observable, and one can unambiguously distinguish two final meso-states for the bead position: the particle either stays at the $F$ trap ($F$ trajectories) or moves with the $M$ trap ($M$ trajectories). In the top panel of Fig.~\ref{figcycle}, we show the ensemble averages  of the position of the bead calculated over $F$ (blue curve) and $M$ (red curve) trajectories.

\begin{figure*}
\includegraphics[width=7 cm]{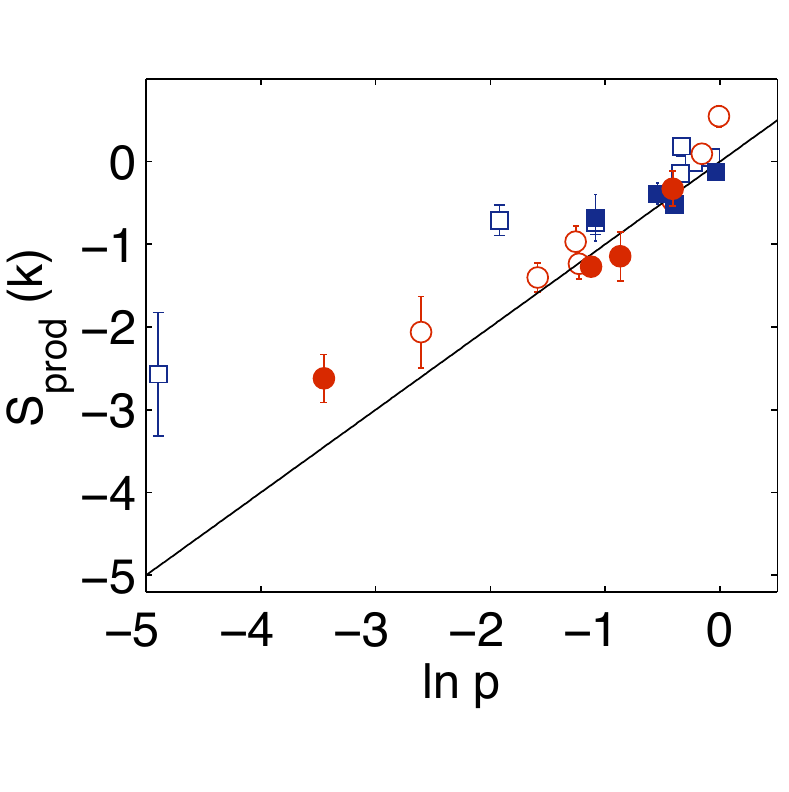}\quad\quad
\includegraphics[width=7 cm]{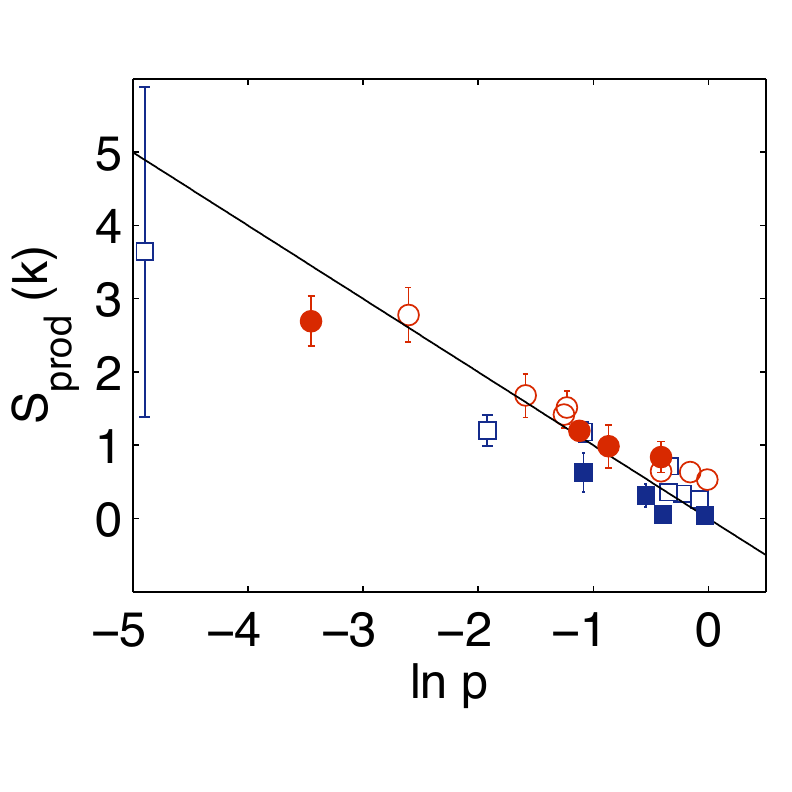}
\caption{Ensemble average conformational entropy production in the SB, $\langle S_{\rm prod}\rangle^{\text{(SB)}}_i$ (left, $k$ units) and in the SR, $\langle S_{\rm prod}\rangle^{\text{(SR)}}_i$ (right, $k$ units) as  functions of the probability $p_i$ ($\tilde p_i$) of adopting instance $i=F,M$. Results shown in open symbols were obtained using the fast protocol ($\tau_2=2\, \rm s$), and results shown in filled symbols were obtained using the slow protocol ($\tau_2=5.5 \,\rm s$). Blue squares represent the ensemble averages over $F$ trajectories, and red circles represent the averages over $M$ trajectories.  Error bars have only statistical sense and were obtained using a statistical significance of $90\%$.} 
\label{slogp}
\end{figure*}

The potential $U (x,t)$ along the protocol (bottom panel in Fig.~\ref{figcycle}) was obtained from the empirical probability density function  calculated combining data from both the SB and the SR. From this potential, we were able to measure the heat or energy transfer from the thermal reservoir to the Brownian particle for individual trajectories \cite{Sekimoto,Seifert2012} and for different values of the external force and therefore of the probability of choice $p_i$ (see Methods). The average conformational entropy production over the $M$ and $F$ realizations for the SB and SR is calculated from the heat and the Shannon entropy, using Eq.~\eqref{mainS}, and plotted in Fig. \ref{slogp} as a function of $\ln p_i$ for the SB and $\ln \tilde p_i$ for the SR. These figures are the main result of the paper. The experiment confirms the dependence of the entropy on the probability of adopting a given instance given by Eq.~\eqref{mainS}.  In the case of the SB, the negative conformational entropy production is clearly observed and the theoretical dependence is reproduced, except for very low probabilities  $p_i\lesssim e^{-2}\simeq 0.05$. We have included 
 error bars calculated using the statistical dispersion of the heat over a large number of cycles. The error in the empirical potential $U(x,t)$ and in the Shannon entropy of the initial and final states, however,  have not been taken into account and could be significant  for small $p_i$, since the number of data points is low. This lack of statistics could explain the discrepancy between the experimental result and the theoretical prediction. The results corresponding to the slow (filled symbols) and the fast (open symbols) protocol are almost indistinguishable, confirming that the quasistatic limit is indeed achieved for the velocities used in the experiment. In the Supplementary Information we have also included numerical simulations of the SB for non-quasistatic processes, to characterize how the dissipative work approaches $kT\ln p_i$ in \eqref{dissipativework}, when the total duration $\tau$ of the process increases.

\section{Building a Szilard engine}

As an illustration of the implications of the previous results, we  construct a Szilard engine that extracts energy from a single thermal reservoir, combining the protocols described above. 
The engine can be implemented with an adequate combination of SB and SR processes where 
the lower bound for the minimal work in Eq.~\eqref{work} is negative. The minimum is attained for $p_i=1/2$ and $\tilde p^i_i=1$, $i=F,M$, as in the original Szilard cycle, but negative work can be achieved for different values of $p_i$ and $\tilde p^i_i$.
We have performed multiple experiments at different conditions and in three experiments
we could achieve a combination of probabilities that gave us a negative average work: 1) $p_F=0.35$, $p_M=0.65$;
2) $\tilde p_M=0.99$; and 3) $\tilde p_F=0.93$. Then, our Szilard engine consists of the following feedback protocol. We start
with the external voltage $V_0$ that gave us the first combination ($p_F=0.35$, $p_M=0.65$) and  measure 
the bead position after the SB. If the bead is in the fixed trap (blue curves in Fig. 3) we change the external field
to the value $V_F$ corresponding to $\tilde p_F=0.93$ and continue the protocol at this value of voltage until the
SR is completed. If after the SB the bead is in the moving  trap (red curves in Fig. 3), we change the external field to
the value $V_M$ that gave us $\tilde p_M=0.99$ and continue the protocol at this value of voltage until the SR
is completed. Finally, the cycle is to be completed by quasistatically tuning the external voltage back to its initial value $V_0$
\footnote{The actual value of the voltages are sensitive to many factors: the construction of the fluid chamber, the optical adjustment of the system, the size of the focal spot of the trap, the chemical composition of the sphere, the output parameters of the generator, etc. For the specific chamber and setup of our experiment, the values were $V_0=2$ V, $V_M=4$ V, and $V_F=0$ V.
}. 
This last step has not been implemented in the experiment, but  in principle it can be realized with arbitrarily small entropy production. 

\begin{figure}
\includegraphics[width=7 cm]{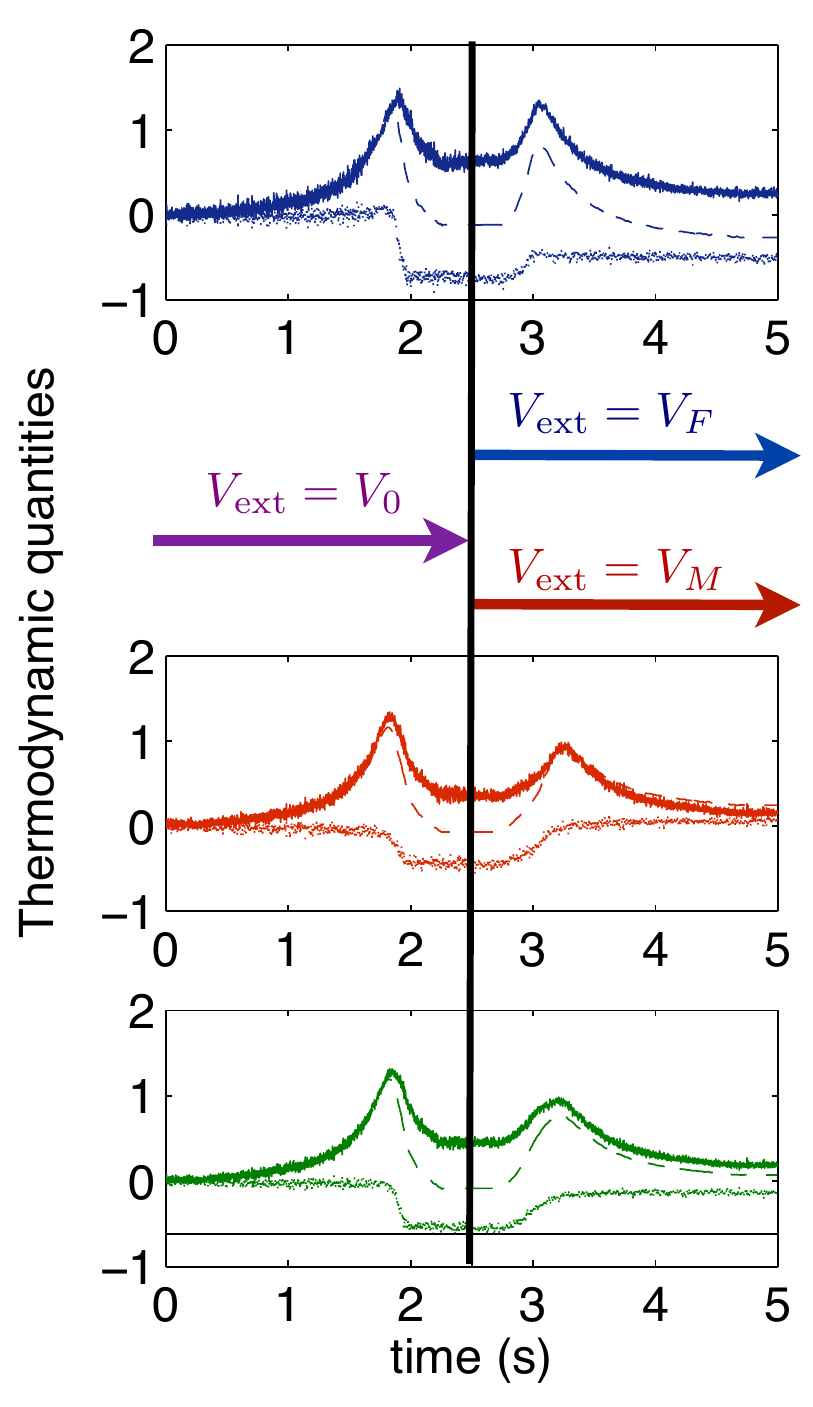}
\caption{Experimental realization of the Szilard engine. We show the average heat (solid lines, $kT$ units), the Shannon entropy of the trajectory distribution (dashed lines, $k$ units), and the average entropy production (dotted lines, $k$ units) as functions of time. The upper plot (blue curves) corresponds to averages over  trajectories that end in the fixed trap, the middle plot (red curves) to averages over trajectories that end in the moving trap, and the lower plot shows averages over all trajectories. The feedback protocol is indicated by the arrows. The symmetry breaking is created with an external voltage $V_{\rm ext}= V_0$ which induces probabilities $p_F=0.35$, $p_M=0.65$. When the particle chooses the fixed trap (blue) the symmetry is restored changing the voltage to $V_{\rm ext}= V_F$, and so biasing the potential towards the fixed trap ($\tilde p_F=0.93$). When the particle ends in the moving trap (red), the symmetry is restored at a voltage $V_{\rm ext}= V_M$, biasing the potential towards the moving trap ($\tilde p_M=0.99$). We also indicate in the bottom figure the value of the relative entropy $D(p_i||\tilde p_i)=\sum_{i}\, p_i\,\ln(p_i/\tilde p_i)$ (black solid line).}
\label{fig:szilard}
\end{figure}

Figure~\ref{fig:szilard} shows the average heat (solid curves), the change in Shannon entropy of the probability distribution of the bead position (dashed curves), and the average entropy production (dotted curves) along the feedback cycle. The averages are taken over trajectories that end in the $F$ trap (upper plot; blue curves), the $M$ trap (middle plot; red curves), and over all trajectories (lower plot; green curves). Notice that the average of the thermodynamic parameters over all trajectories is taken using the probabilities in the SB, that is, $\langle S_{\rm prod} \rangle ^{\text{(SB)}} = \sum_i p_i \langle S_{\rm prod}\rangle^{\text{(SB)}}_i$  for the SB and $\langle S_{\rm prod} \rangle ^{\text{(SR)}} = \sum_i p_i \langle S_{\rm prod}\rangle^{\text{(SR)}}_i$ for the SR. The entropy produced in the whole cycle, averaged over all trajectories, $\langle S_{\rm prod} \rangle = \langle S_{\rm prod} \rangle^{\text{(SB)}} + \langle S_{\rm prod} \rangle^{\text{(SR)}}$,  is negative, as shown in the lower plot in Fig.~\ref{fig:szilard}. Notice that despite being negative, the average entropy production along the cycle is greater than $k\,D(p_i||\tilde{p}_i)=k\sum_i \,p_i\,\ln(p_i/\tilde{p}_i)$ (and greater than the minimum entropy that can be produced $\langle S_{\rm prod} \rangle / k > D(p_i||\tilde{p}_i) > -H(p_i)$~\cite{sagawa2009minimal,horowitz2010nonequilibrium,horowitz2011thermodynamic,horowitz2011iopscience}), as predicted by  Eq.~\eqref{work}.

\section{Conclusions}

Our experiments show  that the signature of a symmetry breaking in the energetics of a  quasistatic process is observable. This signature is relevant in two situations:  estimating the free energy of kinetic or metastable states in macromolecules, and the thermodynamics of computation and information processing. In the first case, the energetics can be used to detect the coexistence of otherwise hidden metastable states, and to identify factors that bias the SB towards a given metastable state. In the second case, the dependence of the entropy production on the probability of adopting a given instance at the SB is able to explain in simple terms the energetics of erasure (Landauer's principle) and feedback (Szilard engine). SB is in fact implicit in other models of information motors  based on memories with  metastable states \cite{mandal2012work} or separation of time scales \cite{horowitz2012imitating}.

Moreover, this signature is universal: it does not depend on the nature of the physical system or the mechanism inducing the SB. It is rather small for SBs involving a limited number of metastable states. However, it could have implications in the way we assess the entropy of systems which have undergone a SB with a large number of instances (or a continuous SB). For instance, biological evolution can be considered as a succession of SBs, where specific sequences of DNA were selected over a gigantic number of possibilities. The same can be said about nucleogenesis in the early universe and other fundamental processes. The conformational entropy in both cases could have experienced a significant decrease, as indicated by Eq. \eqref{mainS}, whose consequences have not yet been explored.

\section{Methods}

\textbf{Experimental setup.}  Polystyrene microspheres of diameter $1\, \mu m$ (G. Kisker-Products for Biotechnology) were diluted in distilled de-ionized water to a final concentration of a few spheres per ml. The spheres were inserted into a custom made electrophoretic fluid chamber with two electrodes connected to a computer controlled electric generator and an amplifier. A $1060\,{\rm nm}$ optical beam is deflected by an acousto-optical deflector AOD  (ISOMET LS55 NIR), expanded and inserted through an oil-immersed objective O1 (Nikon, CFI PL FL 100X NA 1.30) into the fluid chamber. An additional $532 \,{\rm nm}$ optical beam from a laser coupled to a single-mode fiber (OZOptics) is collimated by a $(\times10,\, \text{NA}=0.10)$ microscope objective and passed through the trapping objective. The forward scattered detection beam is collected by a $(\times10,\, \text{NA}=0.10)$ microscope objective O2, and its back focal-plane field distribution is analyzed by a quadrant position detector (QPD) (New Focus 2911) at an acquisition rate of $1\,{\rm kHz}$. The calibrations of the experimental setup are described in the Supplementary Information.

The individual traps (the fixed trap $F$ and the moving trap $M$) are generated with a single beam following a time-sharing protocol for the AOD. The alternation of trap positions is controlled by timing signals generated by a modulation generator adjusted to give a high frequency (20 kHz) square wave with a controllable duty ratio. The time-sharing protocol permitted controlling the position and velocity of the $M$ trap. 

The cycle shown in Fig.~\ref{figcycle} containing a SB and a SR was repeated with the same bead and  electrostatic field  for at least $2400\,\rm s$, corresponding to 480 cycles for the fast and 200 cycles for the slow protocol.
 
\textbf{Data analysis.} We inferred the potential generated by the traps and the external field as a function of time, $U(x,t)$, from the bead position histograms.  The only data used in the analysis comes from the trajectory of the bead along a number of cycles. Since our protocol is quasistatic, we can use as an estimate of the equilibrium probability $\rho_{\rm eq}(x,\lambda(t))$ the empirical PDF of the position of the bead at time $t$. To improve the statistics and also as an extra check of the consistency of this estimate, we combine data from the SB and the SR processes corresponding to the same value of the external parameter $\lambda$. More precisely, we estimate $\rho_{\rm eq}(x,\lambda(t))$Ê as the PDF of the  trajectories of the bead inside two time windows $[t-S/2,t+S/2]$ and $[\tau-t-S/2,\tau-t+S/2]$, of width $S=25 \,{\rm ms}$ and centered at times $t$ and $\tau-t$, respectively, when both traps are at the same position in the steps 2 and 4 of the protocol (see Fig.~\ref{figcycle}). During steps 1 and 3,  the  potential is constant and therefore we can use data from the whole duration of those steps. The bin size used for the PDFs is  $\Delta x=10\,{\rm nm}$.  Using these empirical PDFs we can estimate  $\hat U(x,t)= U(x,t)-kT\ln Z(\lambda(t))=U(x,t)+F(\lambda(t))$ as $-kT \ln \rho_{\rm eq} (x,\lambda(t))$. Note that the heat dissipated to the thermal bath only depends on local properties of the potential. Consequently, the free energy $F(\lambda(t))$ does not enter into the calculation of the production of conformational entropy. We fit our estimate of the potential to a quartic polynomial $\hat U(x,t)=a_0 (t) + a_1(t) x+ a_2(t) x^2 + a_3 (t) x^3 + a_4(t) x^4$ $-$ where $a_i(t)$ are time dependent parameters $-$ using a nonlinear least squares weighted fit every $\Delta t = 1\; \rm ms$. The data were weighted by $w(x,t)=e^{-U(x,t)/kT}$, i.e., we favored the data near the bottom of the wells. The data points that exceed the global minimum of the potential by more than $10kT$ were not considered. 

The conformational entropy production associated to a single stochastic trajectory in the SB has two contributions: the change in the entropy of the particle $\Delta S$ and the entropy flow to the thermal reservoir $-Q/T$:  
\begin{equation}
S_{{\rm prod},i} = \Delta S_i - Q/T.
\end{equation} 
The change in conformational entropy  of the particle is  $\Delta S_i = S_{{\rm fin},i} - S_{\rm ini}$. The initial entropy of the system is given by $S_{\rm ini} = - k \int_{\Gamma} dx \;\rho_{\rm eq}(x,\lambda( t_{\rm ini})) \ln \rho_{\rm eq} (x,\lambda( t_{\rm ini}))$, where the integration is carried out over the whole phase space.  On the other hand, at the end of the process the conformational system entropy depends on the path taken by the bead, $S_{\text{fin, i}} = - k \int_{\Gamma_i} dx \;{\rho}_{\text{eq},i}(x,\lambda(t_{\rm fin}) )\ln \rho_{\text{eq,}i} (x,\lambda( t_{\rm fin}))$, where $\Gamma_i$ ($i=M,F$) is the phase space accessible to the bead after the SB, depending on the path. To obtain $\Delta S_i$ from the experimental trajectories, we use the empirical PDF of the position of the bead using data from the intervals $[0,\tau_1]$ for $\rho_{\rm eq}(x,\lambda( t_{\rm ini}))$ and $[\tau_1+\tau_2,\tau_1+\tau_2+\tau_3]$ for ${\rho}_{\text{eq},i}(x,\lambda(t_{\rm fin}))$.  To estimate the dissipated heat $Q$, we use a modified version of stochastic thermodynamics \cite{Sekimoto,Seifert2012,blickle2011realization}, suitable for potentials that change by small discrete steps (see Supplementary Information).

\section{Acknowledgements}
IM and DP acknowledge financial support from the Fundaci\'o Privada Cellex Barcelona, Generalitat de Catalunya grant 2009-SGR-159, and from the Spanish Ministry of Science and Innovation (MICINN FIS2008-00114, FIS2011-24409). ER and JMRP acknowledge fruitful discussions with Stephan Grill, Marcus Jahnel, Martin Behrndt, Jordan M. Horowitz, and Luis Din\'is, and financial support from grants MOSAICO, ENFASIS (the Government of Spain), and MODELICO (the Comunidad de Madrid).

\end{document}